\documentclass[pra,amsmath,amssymb,preprint ]{revtex4-2}

\usepackage{graphicx}% Include figure files
\usepackage{dcolumn}% Align table columns on decimal point
\usepackage{bm}% bold math
\usepackage[utf8]{inputenc}
\usepackage[T1]{fontenc}
\usepackage{mathptmx}
\usepackage{color}

\begin{document}

\title{New approach to describe two coupled spins in a variable magnetic field}

\author{Yury Belousov}%
 \email[Corresponding author: ]{theorphys@phystech.edu}

 \affiliation{
  Moscow Institute of Physics and Technology, 141700, Institutskii
	Per.\ 9, Dolgoprudny, Moscow Distr., Russian Federation.}
  \affiliation{
  Terra Quantum AG, St.\,Gallerstrasse 16A, 9400 Rorschach, Switzerland.
  }

\author{Roberto Grimaudo}
 \affiliation{
 Dipartimento di Fisica e Chimica dell'Universit\`a di Palermo, Via Archirafi 36, I-90123 Palermo, Italy
  }
\author{Antonino Messina}%
 \affiliation{
  Dipartimento di Matematica ed Informatica dell'Universit\`a di Palermo, Via Archirafi 34, I-90123 Palermo, Italy
  }
  \author{Agostino Migliore}%
 \affiliation{
Dipartimento di Scienze Chimiche, Universit\`a di Padova, 35131 Padova, Italy
% Dipartimento di Scienze Chimiche, Universit\`a di Padova, Via F. Marzolo 1, Padova, Italy
  }
  \author{Alessandro Sergi}%
 \affiliation{
 Dipartimento di Scienze Matematiche e Informatiche, Scienze Fisiche e Scienze della Terra, Universit\`a degli Studi di Messina, viale F. Stagno d'Alcontres 31, 98166 Messina, Italy
 }
 \affiliation{
 Istituto Nazionale di Fisica Nucleare, Sez. di Catania, 95123 Catania, Italy
  }
  \affiliation{
 Institute of Systems Science, Durban University of Technology, P. O. Box 1334, Durban 4000, South Africa
 }

\date{\today} % It is always \today, today, but any date may be explicitly specified
              % Not printed for conference proceedings

\begin{abstract}
We propose a method to describe the evolution of two spins coupled by hyperfine interaction in an external time-dependent magnetic field. We apply the approach to the case of hyperfine interaction with axial symmetry, which can be solved exactly in a constant, appropriately oriented magnetic field. In order to treat the nonstationary dynamical problem, we modify the time-dependent Schr\"odinger equation through a change of representation that, by exploiting an instantaneous (adiabatic) basis makes the time-dependent Hamiltonian diagonal at any time instant. The solution of the transformed time-dependent Schr\"odinger in the form of chronologically ordered exponents with transparent pre-exponential coefficients is reported. This solution is highly simplified when an adiabatically varying magnetic field perturbs the system. The approach here proposed may be used for the perturbative treatment of other dynamical problems with no exact solution.
\end{abstract}

\maketitle

\section{Introduction}

Spin systems are subject of growing interest because of their potential applications in quantum information and pertinent technologies \cite{SB,GW,http}. Two-level spin s =1/2 systems are the classical prototype qubit, namely, the basic unit of quantum information. The behaviors of spin systems influenced by hyperfine interactions have been widely examined because of their relevance to phenomena such as, e.g., NMR and ESR, and related spectroscopic techniques. The great potential for using spin systems in quantum information and computation has opened a broad range of unsolved dynamical problems not previously considered in standard NMR and ESR applications. Adiabatic gates based on the well-known adiabatic approximation have important roles in quantum information processing \cite{A,B,C,MM,AM}.

Within this general context, we here examine the time evolution of two spin systems with hyperfine interaction (HFI) in an adiabatically varying external magnetic field. The Hamiltonian may be generally written
\begin{equation}\label{h1}
   \widehat{H}_{\rm HFI}= \hbar A_{ik} \sigma_{1,i}\sigma_{2,k} + B_i(t)(\gamma^{(1)}_{ik}\sigma_{1,k} +\gamma^{(2)}_{ik}\sigma_{2,k}),
\end{equation}
where $A_{ik}$ is the HFI tensor, $\sigma_{1,i}$ and $\sigma_{2,k}$ are the Pauli operators of the two interacting spins, $\gamma^{(1)}_{ik}$ and $\gamma^{(2)}_{ik}$ are their gyromagnetic ratios, and  $B_i(t)$ is the external magnetic field. The magnetic field is usually described as the sum of a constant and homogeneous field component and a perpendicular alternate (radiofrequency) component: ${\bf B}(t)= {\bf B}_0+{\bf b}(t)$.

Even for the relatively simple system of two interacting spins, the class of dynamical problems defined by Eq. \eqref{h1} does not have a general exact solution, while specializations of Eq. \eqref{h1} have been studied using perturbation theory (see, e.g. \cite{LL}). In the case of a paramagnetic center in a crystal lattice or of a paramagnetic radical, the two spins are not equivalent (see, e.g. \cite{AB}). For a paramagnetic center in a cubic environment or a free radical, the HFI and gyromagnetic ratios are isotropic and the Hamiltonian model has the well-known exact solution used to describe the hyperfine splitting in the hydrogen atom. No exact analytical solution is available for other cases. Nonetheless, the relatively common situation of axially symmetric HFI between the two spins in a static magnetic field can be solved analytically, to a good approximation, using perturbation theory. The solution is based on the technique of approximate diagonalization of Hermitian matrices \cite{B1,B2}. We will extend this technique to study the interacting spins in a time-dependent magnetic field. The solution of the nonstationary case requires finding an evolution operator. This operator has a clear form in a basis of eigenstates of the system Hamiltonian, but the basis vectors will depend on time. The scope of this paper is to show that the knowledge of the time evolution operator of the system under an external (classical) time-independent magnetic field may be suitably exploited to determine the system quantum dynamics under time-dependent magnetic fields. To this end, we will make use of a similar approach reported in refs. \citep{Bagrov,Kuna,DasSarma,MN,GMN,MGMN,GdCNM}.

The approach may be extended to the study of two equivalent spins with a magnetic dipole-dipole interaction. In fact, the HFI also has axial symmetry in these systems, which are of interest for applications in quantum computers and other quantum technologies. Compared to paramagnetic centers, distinctive features of such systems emerge from the equivalent spins and equal magnetic moments of the interacting particles.

\section{Time-dependent Hamiltonian with anisotropic hyperfine interaction}

The Hamiltonian model describes situations in which the symmetry axis of the HFI, $\hat{C}\|{\bf n}$ , is not parallel to the external magnetic field ${\bf B}$. The HFI interaction tensor is written as $A_{ik}=A_i\delta_{ik}$ with respect to the principal axes, and it is determined by the two constants $A_{\|}=A_{zz}$ and $A_{xx}=A_{yy}=A_{\perp}$. We assume g-factors $g^{(a)}_{ik}=g_a\delta_{ik}$ for the two 1/2 spins, with  $g_1 \neq g_2$. Then, $\omega_1 =
g_1\mu_{0,1}B(t) \equiv \omega$ and $\omega_2 = g_2\mu_{0,2}B(t)$. We introduce the ratio $g_2\mu_{0,2}/g_1\mu_{0,1} =\zeta$. As usual in ESR and NMR problems, we express the Hamiltonian in frequency units, by taking $\hbar =1$:
\begin{equation}\label{h2}
 \widehat{H}= \frac{1}{2}\omega(t)\hat{\sigma}_{1z}+
 \frac{\zeta}{2}\omega(t)\hat{\sigma}_{2z}+
 A_{\perp}({\boldsymbol \sigma}_1{\boldsymbol \sigma}_2) +
 (A_{\|}-A_{\perp})({\boldsymbol \sigma}_1{\bf n})({\boldsymbol \sigma}_2{\bf n})
\end{equation}

This is the Hamiltonian model generally used for different paramagnetic centers in crystals. If spin 1 corresponds to an electron and spin 2 to a nucleus, then $g_1 \approx -2$ and $|\zeta|\ll 1$. An additional small parameter can therefore be exploited for perturbative approximations. Another useful parameter is given by the ratio $\omega/A_{\perp, \|} \ll 1$ for low magnetic field or by $A_{\perp, \|}/\omega \ll 1$ in the high-magnetic field limit. These two limits have been studied experimentally \cite{AB}. The Hamiltonian model of Eq.\eqref{h2} does not have an analytical solution of general validity even in the case of constant external magnetic field, for which exact solutions have only been obtained in special cases. The nonstationary dynamical problem needs all the more to be solved using the perturbative approximation. In this study, we will construct a solution of the nonstationary dynamical problem in special cases that have an exact solution in stationary conditions. This aims to be a first step towards a general perturbative solution of the nonstationary problem.

The article is organized as follows. First, we examine a class of stationary situations in which the dynamical problem can be solved exactly. Then, we exploit the results of the stationary case to derive a convenient form of the Schr\"odinger equation in the nonstationary case. Finally, we find the solution in the nonstationary conditions.

\section{\label{sec:S}Stationary case}

When the external magnetic field is a constant, and thus $\omega(t) \equiv \omega$, a solution of the Hamiltonian model for special field orientations of interest can be found through a technique consisting of the approximate diagonalization of the Hamiltonian matrix. This technique was successfully used in the study of $\mu$SR in semiconductors to describe the behavior of the muon spin polarization (e.g., see \cite{B1,B2}).

We represent the two-spin state in the standard basis set
\begin{equation}\label{h3}
  |\chi_1\rangle = |+\rangle|+\rangle,\, |\chi_2\rangle = |+\rangle|-\rangle,\,
  |\chi_3\rangle = |-\rangle|+\rangle,\, |\chi_4\rangle = |-\rangle|-\rangle,
\end{equation}
where the first and second vectors describe the projections of the electronic and nuclear spins on the $z$ axis, respectively. Here, we consider the two cases solved in stationary conditions, in which the magnetic field is parallel, ${\bf B} \|{\bf n}$ ($\theta =0$), and orthogonal, ${\bf B} \perp{\bf n}$  ($\theta =\pi/2$), to the symmetry axis of the two-spin system.

Using basis set \eqref{h3}, the Hamiltonian matrix for the ${\bf B}\|{\bf n}$ case reads
\begin{equation}\label{h4}
  \widehat{H}\Bigr |_{\theta=0}=
  \left(
  \begin{array}{llll}
    A_{\|}+\omega(1+\zeta)/2 & 0 &0 & 0 \\
    0 & -A_{\|}+\omega(1-\zeta)/2 &2A_{\perp} & 0 \\
    0 & 2A_{\perp} & -A_{\|}-\omega(1-\zeta)/2 & 0 \\
    0 & 0 & 0 & A_{\|}-\omega(1+\zeta)/2
  \end{array}
  \right),
\end{equation}
and its eigenvalues are
\begin{equation}\label{h5}
    \varepsilon_{1,4}=A_{\|}\pm\omega(1+\zeta)/2, \quad
    \varepsilon_{2,3}=-A_{\|}\pm \frac{1}{2}\sqrt{4A_{\perp}^2 +\omega^2(1-\zeta)^2}.
\end{equation}

In the case $\theta = \pi/2$, we have
\begin{equation}\label{h6}
  \widehat{H}\Bigr |_{\theta=\pi/2}=
  \left(
  \begin{array}{llll}
    A_{\perp}+\omega(1+\zeta)/2 & 0 &0 & A_{\|}-A_{\perp} \\
    0 & -A_{\perp}+\omega(1-\zeta)/2 &A_{\|}+A_{\perp} & 0 \\
    0 & A_{\|}+A_{\perp} & -A_{\perp}-\omega(1-\zeta)/2 & 0 \\
    A_{\|}-A_{\perp}  & 0 & 0 & A_{\perp}-\omega(1+\zeta)/2
  \end{array}
  \right),
\end{equation}
with energy eigenvalues
\begin{equation}\label{h7}
    \varepsilon_{1,4}=A_{\perp}\pm \frac{1}{2} \sqrt{4(A_{\perp}-A_{\|})^2+\omega^2(1+\zeta)^2}, \quad
  \varepsilon_{2,3}=-A_{\perp}\pm \frac{1}{2} \sqrt{4(A_{\perp}+A_{\|})^2+\omega^2(1-\zeta)^2}.
\end{equation}

Solutions \eqref{h5} and \eqref{h7} are easily found by transforming the basis of Eq. \eqref{h3}  through the unitary operation
\begin{equation}\label{h8}
  T=
  \left(
  \begin{array}{cccc}
    \cos\vartheta_2 & 0& 0& -\sin\vartheta_2  \\
    0 & \cos\vartheta_1 & -\sin\vartheta_1 &0\\
    0 & \sin\vartheta_1 &\cos\vartheta_1 &0\\
   \sin\vartheta_2 & 0 & 0 & \cos\vartheta_2
  \end{array}
  \right),
\end{equation}
where
\[
 \tan(2\vartheta_1)=\frac{2(2A_{\perp}+\Delta A)}{\omega(1-\zeta)}; \quad
 \tan(2\vartheta_2)=\frac{2\Delta A}{\omega(1+\zeta)} \quad \mbox{and} \quad
 \Delta A=(A_{\|}-A_{\perp})\sin^2\theta.
\]
It is $\vartheta_2=0$ for ${\bf B}\|{\bf n}$ at any time. In the absence of magnetic field,  $\vartheta_1=\pi/2$. In the limit $B \to 0$ for ${\bf B}\perp {\bf n}$ we obtain $\vartheta_1=\pi/2$. The unitary transformation of Eq.\eqref{h8} produces the diagonal matrix $\widehat{\widetilde{H}}= T(\theta) \widehat{H} T^{\dag}(\theta)$, with diagonal elements given by Eq.s \eqref{h5} and \eqref{h7}. The corresponding eigenstates are easily obtained as superpositions of the basis vectors.

\section{Non-stationary case}

As in the stationary case, the dynamical problem is reduced to a two-level problem when the nonstationary magnetic field is oriented parallel or orthogonal to the symmetry axis of the system. Since the evolution operator can always be found for a two-level system \cite{MN}, the comparison between exact and approximate solutions can validate the quality of the here proposed approximation technique and help finding a solution to the general dynamical problem (for arbitrary orientations of the external field) in future studies.

The state of the system at time $t$ is given by
\begin{equation}\label{h9}
    |\chi(t)\rangle  = U(t,0) |\chi(0)\rangle,
\end{equation}
where $\widehat{U}(t,0)$ is the evolution operator and $|\chi(0)\rangle$ is the state of the system at the initial time.

The time evolution operator $\widehat{U}(t)$, which is solution of the Schr\"odinger equation  $i\dot{U}=HU$, can be easily found by noting that, for both $\theta = 0$ and $\theta = pi/2$, the two-spin dynamical problem can be reduced to two single-spin dynamical problems \cite{MN}. In the first case, $\theta = 0$, we only have a relevant two-level sub-dynamics (corresponding to the matrix central block), and the time evolution operator is written
\begin{equation}\label{h10}
U=
\begin{pmatrix}
\exp\{-i(A_{\|}t+(1+\zeta)/2\int_0^t\omega(t') dt')\} & 0 & 0 & 0 \\
0 & \alpha & \beta & 0 \\
0 & -\beta^* & \alpha^* & 0 \\
0 & 0 & 0 & \exp\{-i(A_{\|}t-(1+\zeta)/2\int_0^t\omega(t') dt')\}
\end{pmatrix}
\end{equation}
In the second case, $\theta = pi/2$, we obtain two single-spin sub-dynamics and the time evolution operator reads
\begin{equation}\label{h11}
U=
\begin{pmatrix}
\alpha _2 & 0 & 0 & \beta_2 \\
0 & \alpha _1 & \beta_1 & 0 \\
0 & -\beta_1^* & \alpha _1^* & 0 \\
-\beta_2^* & 0 & 0 & \alpha _2^*
\end{pmatrix}
\end{equation}

It is worth noting that the explicit expressions of parameters  $\alpha$ and $\beta$ depend on the specific time dependence of the magnetic field, that is, on the form of $\omega(t)$. It is generally difficult to find analytical expressions for these two parameters, but such expressions were found for the Landau-Majorana-St\"uckelberg-Zener dynamics \cite{LMSZ} and the Rabi dynamics \cite{Rabi,Rabi2}.

The Schr\"odinger equation reads
\begin{equation}\label{h12}
  i\frac{\partial}{\partial t}|\chi(t)\rangle =
  \widehat{H}(t)|\chi(t)\rangle.
\end{equation}
where the time-dependent Hamiltonian is given by Eq. \eqref{h2}. Since an evolution operator can be found when the Hamiltonian matrix has the form of Eq. \eqref{h4} or \eqref{h6}, we carry out an ``instantaneous'' transformation as in Eq. \eqref{h8} to diagonalize the Hamiltonian matrices \eqref{h4} and \eqref{h6} at each time $t$. That is, the time-dependent unitary transformation has the same form as in Eq. \eqref{h8}, but with time-dependent matrix elements. It is worth noting that, while the procedure can be applied independently of the rate of field change, and thus independently of the applicability of the adiabatic approximation, the time-dependent eigenstates and energy eigenvalues retain a clear physical meaning, as evolutes of the initial ones, only when the adiabatic approximation is applicable. That is, if the system is prepared in an eigenstate of the Hamiltonian at the initial time, the system remains in the evolved eigenstate of the Hamiltonian at time $t$ only if the variation of the magnetic field is sufficiently slow to satisfy the adiabatic approximation, i.e.,
\begin{equation}\label{h13}
  \frac{\dot{\omega}(t)}{\omega^2(t)} \ll 1, \quad \mbox{where}
  \quad \dot{\omega}(t)=\frac{d}{d t}\omega(t).
\end{equation}

We introduce a 'quasi-interaction representation' similarly to the standard interaction representation, but using the unitary evolution operator of Eq. \eqref{h9} (and we define it as an adiabatic representation):
\begin{equation}\label{h14}
  |\chi(t)\rangle = \widehat{T}(t)|\varphi(t)\rangle, \quad
  |\varphi(0)\rangle=\widehat{T}^{\dag}(0)|\chi(0)\rangle,
\end{equation}
The second Eq.\eqref{h14} is necessary because  $\widehat{T}(0)\neq \hat 1$. The insertion of this transformation into Eq. \eqref{h12} gives
\[
  i{\dot{\widehat T}}(t)|\varphi(t)\rangle +
  i\widehat{T}(t)|\frac{\partial}{\partial t}\varphi(t)\rangle=
  \widehat{H}\widehat{T}(t)|\varphi(t)\rangle
\]
and, multiplying both sides of this equation by $\widehat{T}^{\dag}(t)$, we obtain
\begin{equation}\label{h15}
 i\frac{\partial}{\partial t}|\varphi(t)\rangle=
  \widehat{T}^{\dag}(t)\widehat{H}\widehat{T}(t)|\varphi(t)\rangle-
  i\widehat{T}^{\dag}(t){\dot{\widehat T}}(t)|\varphi(t)\rangle.
\end{equation}
In Eq. \eqref{h15}, the transformed Hamiltonian $\widehat{T}^{\dag}(t)\widehat{H}\widehat{T}(t)$  results from matrix \eqref{h4} for  $\theta=0$ and from matrix \eqref{h6} for $\theta=\pi|2$. The additional operator on the right-hand side of Eq. \eqref{h15} has the off-diagonal form
\begin{equation}\label{h16}
 \widehat{T}^{\dag}(t){\dot{\widehat T}}(t)=
 \left(
  \begin{array}{cccc}
    0 & 0 &0& -\dot{\vartheta}_2 \\
    0 &0& -\dot{\vartheta}_1 &0 \\
    0& \dot{\vartheta}_1 &0&0 \\
    \dot{\vartheta}_2 &0&0&0
  \end{array}
 \right); \quad
  \widehat{T}^{\dag}(t){\dot{\widehat T}}(t)
  =- {\dot{\widehat T}}^{\dag}(t)\widehat{T}(t)
\end{equation}
Note that the additional term on the right-hand side of Eq. \eqref{h15}, which is described by Eq. \eqref{h16}, does not affect the possibility of reducing the actual two-spin dynamical problem to two independent two-level dynamical problems. Defining $\widehat{T}^{\dag}(t)\widehat{H}\widehat{T}(t)=\widetilde{H}(t)$, we recast eq. \eqref{h15} in the form
\begin{equation}\label{h17}
 i\frac{\partial}{\partial t}|\varphi(t)\rangle=
  \left(\widetilde{\widehat{H}}(t)-
  i\widehat{T}^{\dag}(t){\dot{\widehat T}}(t)\right)|\varphi(t)\rangle.
\end{equation}
which represents the modified Schr\"odinger equation for the transformed state. As only the external field, or $\omega(t)$, depends on time, the time derivatives $\dot{\vartheta}_1$ and
$\dot{\vartheta}_2$ are given by
\begin{equation}\label{h18}
   - \dot{\vartheta}_1=
   \left\{
     \begin{array}{ll}
       \frac{2A_{\perp}(1-\zeta)}{16A_{\perp}^2+\omega^2(1-\zeta)^2}\dot{\omega},
& \hbox{$\theta=0$;} \\
        \frac{(A_{\|}+A_{\perp}(1-\zeta)}{4(A_{\|}+A_{\perp})^2+\omega^2(1-\zeta)^2}\dot{\omega}, & \hbox{$\theta=\pi/2$.}
     \end{array}
   \right. \qquad
- \dot{\vartheta}_2=
   \left\{
     \begin{array}{ll}
       0, & \hbox{$\theta=0$;} \\
        \frac{(A_{\|}-A_{\perp}(1+\zeta)}{4(A_{\|}-A_{\perp})^2+\omega^2(1+\zeta)^2}\dot{\omega}, & \hbox{$\theta=\pi/2$.}
     \end{array}
   \right.
\end{equation}
Next, we analyze the results of our approach for the two exactly solvable cases in which the orientation of the magnetic field is defined by $\theta =0$ and $\pi/2$.

\section{Results and discussions}

\subsection{Magnetic field parallel to the HFI symmetry axis  ($\theta=0$)}

We consider the case $\theta=0$ for an arbitrary time dependence of the magnetic field. As shown by Eq. \eqref{h10}, the evolution operator for states $|\varphi_1\rangle$ and $|\varphi_4\rangle$ is determined by the time dependence of $\varepsilon_{1,4}(t)$ in Eq. \eqref{h5}. The dynamical problem is reduced to the solution of a 2$\times$2 problem with the Hamiltonian matrix
\begin{equation}\label{h19}
    \widehat{H}_{23}(\theta =0)=
 \left(
  \begin{array}{cc}
    \varepsilon_2(t)& -\dot{\vartheta}_1 \\
     \dot{\vartheta}_1 &\varepsilon_3(t)
  \end{array}
 \right)
\end{equation}
in which  $\varepsilon_{2,3}(t)$ are given by Eq. \eqref{h5} and $\dot{\vartheta}_1(0)$ has the expression in Eq. \eqref{h18}.

We examine the evolution of the states $|\varphi(t)\rangle= \widehat{T}^{\dagger}(t)|\chi(t)\rangle$ determined by the Hamiltonian in Eq. \eqref{h19}. Eq. \eqref{h14} gives the initial conditions
\[
  |\varphi_1(0)\rangle = |+\rangle|+\rangle, \quad
  |\varphi_4(0)\rangle = |-\rangle|-\rangle,
\]
and
\begin{equation}\label{h20}
  |\varphi_2(0)\rangle = a_+(\theta=0)|+\rangle|-\rangle  +a_-(\theta=0)|-\rangle|+\rangle, \quad
  |\varphi_3(0)\rangle = -a_-(\theta=0)|+\rangle|-\rangle  +a_+(\theta=0)|-\rangle|+\rangle,
\end{equation}
where
\begin{equation}\label{h21}
  a_{\pm}(\theta=0)=\frac{1}{\sqrt{2}}\sqrt{1\pm \frac{\omega(t=0)(1-\zeta)}{\sqrt{(4A_{\perp})^2+\omega^2(t=0)(1-\zeta)^2}}}.
\end{equation}
We can treat the dynamical problem in the interaction representation by considering the $\dot{\vartheta}_1(t)$ term as a perturbation. The perturbation in Eq. \eqref{h19} has the form
\begin{equation}\label{h22}
    \widehat{V}_{23}(t)=i\frac{A_{\perp}(1-\zeta)\dot{\omega}(t)}{(4A_{\perp})^2    +\omega^2(t)(1-\zeta)^2} \sigma_y.
\end{equation}

In accordance with Eqs. \eqref{h5} and \eqref{h19}, the evolution operator for the unperturbed system is
\begin{equation}\label{h23}
    U_{23}^{(0)}(t)=e^{iA_{\|}t}\exp\left(
    -\frac{i}{2}\int\limits_0^t\sqrt{4A_{\perp}^2+\omega^2(t')(1-\zeta)^2}d t'\sigma_z
    \right).
\end{equation}

In the interaction representation, the operator of Eq. \eqref{h22} becomes
\begin{align}\label{h24}
    \widehat{V}^{I}_{23}(t)=&{U_{23}^{(0)}}^{\dagger}(t)\widehat{V}_{23}(t) U_{23}^{(0)}(t)=
    \nonumber \\
   =& i\frac{A_{\perp}(1-\zeta)\dot{\omega}(t)}{(4A_{\perp})^2+\omega^2(t)(1-\zeta)^2}
    \left(\sigma_y\cos\int_0^t\omega_{23}(t')dt' +
    \sigma_x\sin\int_0^t\omega_{23}(t')dt'\right),
\end{align}
where $\omega_{23}(t)=\sqrt{(4A_{\perp})^2+\omega^2(t)(1-\zeta)^2}$.

Finally, we can write the following expression for states $|\varphi_{2,3}(t)\rangle$:
\begin{equation}\label{h25}
    |\varphi_{2,3}(t)\rangle =e^{iA_{\|}t}\exp\left(
    \mp \frac{i}{2}\int\limits_0^t\sqrt{4A_{\perp}^2+\omega^2(t')(1-\zeta)^2}d t'\right)
    \widehat{T}\exp\left(-i\int\limits_0^t\widehat{V}^{I}_{23}(t')d\,t'\right)|\varphi_{2,3}(0)\rangle,
\end{equation}
where $\widehat{V}^{I}_{23}(t)$ is given by Eq. \eqref{h24}.

Eq. \eqref{h25} provides a formally exact solution of the dynamical problem, and the first-order approximation in the perturbation yields parameters $\alpha$ and $\beta$ in Eq.\eqref{h10}.

\subsection{Magnetic field perpendicular to the HFI symmetry axis ($\theta=\pi/2$)}

In this case, one of the two-level subsystems is a mixture of states $|\varphi_1\rangle$ and $|\varphi_4\rangle$, and the other is a mixture of states $|\varphi_2\rangle$ and $|\varphi_3\rangle$ as in the previous case. The unperturbed energy levels are provided by Eq. \eqref{h7}. We write the two 2$\times$2 Hamiltonian matrices (including the perturbations) as
\begin{equation}\label{h26}
    \widehat{H}_{14}(\theta =\pi/2)=
 \left(
  \begin{array}{cc}
    \varepsilon_1(t)& -\dot{\vartheta}_2 \\
     \dot{\vartheta}_2 &\varepsilon_4(t)
  \end{array}
 \right)
 \quad \mbox{and} \quad
     \widehat{H}_{23}(\theta =\pi/2)=
 \left(
  \begin{array}{cc}
    \varepsilon_2(t)& -\dot{\vartheta}_1 \\
     \dot{\vartheta}_1 &\varepsilon_3(t)
  \end{array}
 \right)
\end{equation}
where  $\varepsilon_{1,4}(t)$ and $\varepsilon_{2,3}(t)$ are obtained by introducing the time-dependence of the frequency in Eqs. \eqref{h7}. The state vectors at time zero, $|\varphi_i(0)\rangle$, are
\begin{align}\label{h27}
  |\varphi_1(0)\rangle = & b_+(\theta=\pi/2)|+\rangle|-\rangle  +b_-(\theta=\pi/2)|-\rangle|+\rangle, \nonumber \\
  |\varphi_4(0)\rangle =& -b_-(\theta=\pi/2)|+\rangle|-\rangle  +b_+(\theta=\pi/2)|-\rangle|+\rangle,
\end{align}
where $b_+(\theta=\pi/2)=\cos\vartheta_2(\theta=\pi/2, t=0), \, b_-(\theta=\pi/2)=\sin\vartheta_1(\theta=\pi/2, t=0)$,
that is,
\begin{equation}\label{h28}
  b_{\pm}(\theta=\pi/2)=\frac{1}{\sqrt{2}}\sqrt{1\pm \frac{\omega(t=0)(1+\zeta)}{\sqrt{4(A_{\|}-A_{\perp})^2+\omega^2(t=0)(1+\zeta)^2}}}.
\end{equation}
Similarly,
\begin{align}\label{h29}
  |\varphi_2(0)\rangle = & a_+(\theta=\pi/2)|+\rangle|-\rangle  +a_-(\theta=\pi/2)|-\rangle|+\rangle, \nonumber \\
  |\varphi_3(0)\rangle = & -a_-(\theta=\pi/2)|+\rangle|-\rangle  +a_+(\theta=\pi/2)|-\rangle|+\rangle,
\end{align}
where $a_+(\theta=\pi/2)=\cos\vartheta_1(\theta=\pi/2, t=0), \, a_-(\theta=\pi/2)=\sin\vartheta_1(\theta=\pi/2, t=0)$, namely,
\begin{equation}\label{h30}
  a_{\pm}(\theta=\pi/2)=\frac{1}{\sqrt{2}}\sqrt{1\pm \frac{\omega(t=0)(1-\zeta)}{\sqrt{4(A_{\|}+A_{\perp})^2+\omega^2(t=0)(1-\zeta)^2}}}.
\end{equation}
The perturbations for the two subsystems have the expressions
\begin{equation}\label{h31}
 \widehat{V}_{14}(t)=i\frac{2(A_{\|}-A_{\perp})(1+\zeta)\dot{\omega}(t)}{4A_{\|}-A_{\perp})^2    +\omega^2(t)(1+\zeta)^2} \sigma_y, \quad
    \widehat{V}_{23}(t)=i\frac{(A_{\|}+A_{\perp})(1-\zeta)\dot{\omega}(t)}{4(A_{\|}+A_{\perp})^2    +\omega^2(t)(1-\zeta)^2} \sigma_y.
\end{equation}
and, using the energy levels in Eq. \eqref{h7}, the evolution operators for the two unperturbed subsystems are
\begin{eqnarray}
% \nonumber to remove numbering (before each equation)
     U_0^{(14)}(t)&=&e^{-iA_{\perp}t}\exp\left(
    -\frac{i}{2}\int\limits_0^t\sqrt{4(A_{\|}-A_{\perp})^2+\omega^2(t')(1+\zeta)^2}d t'\sigma_z
    \right), \label{h32}\\
   U_0^{(23)}(t)&=&e^{iA_{\perp}t}\exp\left(
    -\frac{i}{2}\int\limits_0^t\sqrt{4(A_{\|}+A_{\perp})^2+\omega^2(t')(1-\zeta)^2}d t'\sigma_z
    \right). \label{h33}
\end{eqnarray}

In accordance with Eq. \eqref{h26}, perturbations \eqref{h31} in the interaction representation are written
\begin{align}
    \widehat{V}^{I}_{14}(t)
   =& i\frac{2(A_{\|}-A_{\perp})(1+\zeta)\dot{\omega}(t)}{4(A_{\|}-A_{\perp})^2+\omega^2(t)(1+\zeta)^2}
    \left(\sigma_y\cos\int_0^t\omega_{14}(t')dt' +
    \sigma_x\sin\int_0^t\omega_{14}(t')dt'\right), \label{h34}\\
    \widehat{V}^{I}_{23}(t)
   =& i\frac{(A_{\|}+A_{\perp})(1-\zeta)\dot{\omega}(t)}{4(A_{\|}+A_{\perp})^2+\omega^2(t)(1-\zeta)^2}
    \left(\sigma_y\cos\int_0^t\omega_{23}(t')dt' +
    \sigma_x\sin\int_0^t\omega_{23}(t')dt'\right), \label{h35}
\end{align}
where $\omega_{14}(t)=\sqrt{4(A_{\|}-A_{\perp})^2+\omega^2(t)(1+\zeta)^2}$ and $\omega_{23}(t)=\sqrt{4(A_{\|}+A_{\perp})^2+\omega^2(t)(1-\zeta)^2}$.

Finally, the state vectors for the two subsystems are
\begin{eqnarray}
% \nonumber to remove numbering (before each equation)
  |\varphi_{1,4}(t)\rangle =e^{-iA_{\perp}t}\exp\left(
    \mp \frac{i}{2}\int\limits_0^t\sqrt{4(A_{\|}-A_{\perp})^2+\omega^2(t')(1+\zeta)^2}d t'\right)
    \widehat{T}\exp\left(-i\int\limits_0^t\widehat{V}^{I}_{23}(t')d\,t'\right)|\varphi_{2,3}(0)\rangle, \label{h36} \\
    |\varphi_{2,3}(t)\rangle =e^{iA_{\perp}t}\exp\left(
    \mp \frac{i}{2}\int\limits_0^t\sqrt{4(A_{\|}+A_{\perp})^2+\omega^2(t')(1-\zeta)^2}d t'\right)
    \widehat{T}\exp\left(-i\int\limits_0^t\widehat{V}^{I}_{23}(t')d\,t'\right)|\varphi_{2,3}(0)\rangle. \label{h37}
\end{eqnarray}
which represent an exact solution of the dynamical problem. The parameters $\alpha_{1,2}$ and $\beta_{1,2}$ in \eqref{h11} can be derived from the approximation to the first order in the perturbations in Eqs. \eqref{h34} and \eqref{h35}.

\section{Conclusion}

We showed that the adiabatic representation enables a clear description of the time evolution of spin states. Eqs. \eqref{h25} and \eqref{h36}-\eqref{h37} provide exact solutions to the dynamical problem for the two special cases in which the external time-dependent magnetic field is parallel and perpendicular to the symmetry axis, respectively. These exact expressions involve $T$-exponents whose cumbersome explicit expressions are not reported here. In the adiabatic approximation, all $T$-exponents can be approximated to the first order, thus providing corrections to the first-order in the rate of change of the external field.

\end{document}